# Optical properties of a four-layer waveguiding nanocomposite structure in near-IR regime

I S Panyaev[1], N N Dadoenkova[1,2], Yu S Dadoenkova[1,2,3], I A Rozhleys[4], M Krawczyk[5], I L Lyubchanskii[2], and D G Sannikov[1]

**Abstract** The theoretical study of the optical properties of TE- and TM- modes in a four-layer structure composed of the magneto-optical yttrium iron garnet guiding layer on a dielectric substrate covered by planar nanocomposite guiding multilayer is presented. The dispersion equation is obtained taking into account the bigyrotropic properties of yttrium-iron garnet, and an original algorithm for the guided modes identification is proposed. The dispersion spectra are analyzed and the energy flux distributions across the structure are calculated. The fourfold difference between the partial power fluxes within the waveguide layers is achieved in the wavelength range of 200 nm.

**Keywords** Nanocomposite, Optical waveguide, Long-wave approximation

[1] Ulyanovsk State University, 432017 Ulyanovsk, Russian Federation

[2] Donetsk Physical and Technical Institute of the National Academy of Sciences of Ukraine, 83114 Donetsk, Ukraine

[3] Novgorod State University, 173003 Veliky Novgorod, Russian Federation

[4] National Research Nuclear University MEPhI, 115409 Moscow, Russian Federation

[5] Faculty of Physics, Adam Mickiewicz University in Poznań, 61–614 Poznań, Poland

E-mail: sannikov-dg@yandex.ru

## 1 Introduction

The investigations of the optical four-layer structures began with the early theoretical works on isotropic waveguides (Tien et al. 1969; Tien and Ulrich 1970; Tien et al. 1972). The early experimental measurements of the waveguide modal characteristics have been carried out by Tien *et al*. (Tien et al. 1973) and Sun and Muller (Sun and Muller 1977). The four-layer structures have been used, for example, in the polarizers of TE- and TM-modes (Polky and Mitchell 1974), leaky modes lasers (Scifres et al. 1976), optical waveguide lenses and tapered couplers (Southwell 1977), thin-film Luneburg lens (Hewak and Lit 1987; Tabib-Azar 1995), mid-infrared modulators and switches (Stiens et al. 1997; Stiens et al. 1994), *etc*.

On the other hand, multilayered nanocomposite (NC) structures are of great interest due to their specific properties which make them suitable for wide applications (Haus 2016). During the last decade different multilayered waveguide structures have been studied: the magnetic photonic crystals (Khokhlov et al. 2015; Khosravi et al. 2015; Sylgacheva et al. 2016a; Sylgacheva et al. 2016b), multilayered waveguides (Dotsch et al. 2005), silicon-based hybrid gap surface plasmon polariton waveguides (Rao and Tang 2012), nanophotonic and plasmonic waveguides (Alaeian and Dionne 2014) and multilayer graphene waveguides (Smirnova et al. 2014).

In this paper, the optical properties of TE- and TM-waves in the hybrid four-layer magneto-optical (MO) nanocomposite waveguide are investigated. The structure comprises a MO layer on dielectric substrate covered by multilayered dielectric NC medium, which is described in terms of the effective medium approach (Brekhovskikh 1980; Rytov 1956). The numerical computations of the dispersion and energy dependencies in the near infra-red (IR) regime are performed using the obtained dispersion relation. The Q-factor modulation of the fluxes in the waveguide layers is calculated and some possible applications of the structure are discussed.

## 2 Theoretical analysis

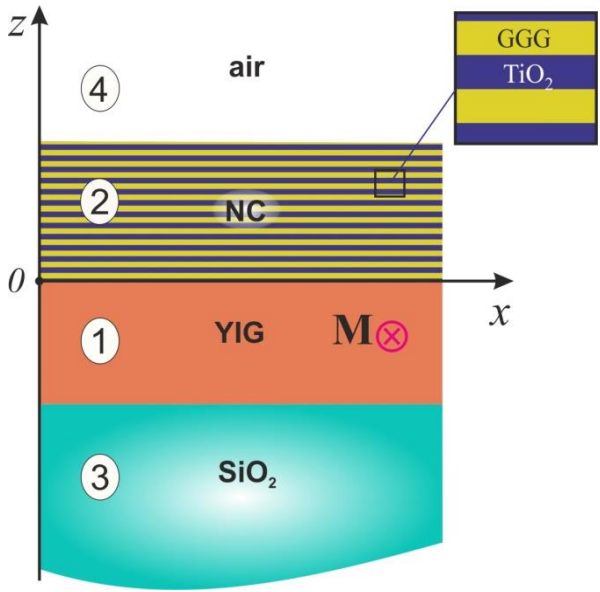

**Fig. 1.** Schematic arrangement of the guiding structure: layer 1 – YIG; layer 2 – NC multilayer; layer 3 – $SiO_2$ substrate; layer 4 – air.

The four-layered waveguide structure consists of a thick $SiO_2$ substrate, a MO YIG film of a thickness $L_1$, covered by the one-dimensional NC multilayer of the thickness $L_2$ and air playing the role of a cladding (as shown in Fig. 1). The NC multilayer with the period $d = d_{GGG} + d_{TiO_2}$ is formed by alternating nanolayers of gadolinium-gallium garnet (GGG) $Gd_3Ga_5O_{12}$ and titanium oxide $TiO_2$ with the corresponding thicknesses $d_{GGG}$ and $d_{TiO_2}$. We consider the transverse MO configuration, so the electromagnetic wave propagates along the $x$-axis, and the magnetization vector **M** is perpendicular to the wave propagation direction and directed along the $y$-axis. In this case the electromagnetic eigenwaves in the YIG layer split into the independent TE- and TM-modes (Zvezdin and Kotov 1997).

It is well known that YIG is transparent in the near IR regime (Randoshkin and Chervonenkis 1990; Zvezdin and Kotov 1997) and exhibits bigyrotropic properties, so both dielectric permittivity $\hat{\varepsilon}_{YIG}$ and magnetic permeability $\hat{\mu}_{YIG}$ tensors contain nonzero off-diagonal components. In the linear MO approximation, for **M** directed along the $y$-axis, the permittivity and permeability tensors have the following form (Gurevich and Melkov 1996):

$$\hat{\varepsilon}_{YIG} = \begin{pmatrix} \varepsilon_1 & 0 & i\varepsilon_a \\ 0 & \varepsilon_1 & 0 \\ -i\varepsilon_a & 0 & \varepsilon_1 \end{pmatrix}, \quad \hat{\mu}_{YIG} = \begin{pmatrix} \mu_1 & 0 & i\mu_a \\ 0 & \mu_1 & 0 \\ -i\mu_a & 0 & \mu_1 \end{pmatrix}. \tag{1}$$

The permittivity tensors of the materials, constituting the NC (GGG and TiO$_2$) are diagonal: $\hat{\varepsilon}_{GGG} = \varepsilon_{GGG}\hat{I}$ and $\hat{\varepsilon}_{TiO_2} = \varepsilon_{TiO_2}\hat{I}$, where $\hat{I} = \delta_{ik}$, and $\delta_{ik}$ is the Kronecker's delta. We assume $d_{GGG}, d_{TiO_2} << \lambda$ ($\lambda$ is the wavelength of the electromagnetic wave), thus the long-wave approximation can be used, so the NC permittivity has the form (Agranovich and Kravtsov 1985):

$$\hat{\varepsilon}_{NC} = \begin{pmatrix} \varepsilon_{xx} & 0 & 0 \\ 0 & \varepsilon_{yy} & 0 \\ 0 & 0 & \varepsilon_{zz} \end{pmatrix}, \quad \varepsilon_{xx} = \varepsilon_{yy} = \frac{\Theta\varepsilon_{GGG} + \varepsilon_{TiO_2}}{\Theta + 1}, \quad \varepsilon_{zz} = \frac{\varepsilon_{GGG}\varepsilon_{TiO_2}(\Theta + 1)}{\Theta\varepsilon_{TiO_2} + \varepsilon_{GGG}}, \tag{2}$$

where $\Theta = d_{GGG}/d_{TiO_2}$ is the ratio of the GGG and TiO$_2$ layers thicknesses.

The substrate (SiO$_2$) and the cladding (air) have the scalar permittivities $\varepsilon_3$, $\varepsilon_4$, respectively. In the considered frequency range (the near IR regime) the permeabilities of each layer are equal to unity, *i.e.* $\mu_1 = \mu_2 = \mu_3 = \mu_4 = 1$.

The electric and magnetic fields of the electromagnetic wave of the angular frequency $\omega$, propagating along the $x$-axis, are proportional to $\exp[i(\omega t - \beta x)]$, where $\beta$ is the propagation constant. The tangential components of the vector profile function $\mathbf{F}(z)$ (the electric field component $E_y$ for TE-mode and the magnetic field component $H_y$ for TM-mode) have the following form:

$$F_y(z) = A \cdot \begin{cases} \cos h_1 z + C_1 \sin h_1 z, & -L_1 \leq z \leq 0, \\ \cos h_2 z + C_2 \sin h_2 z, & 0 \leq z \leq L_2, \\ \left[\cos h_1 L_1 - C_1 \sin h_1 L_1\right] e^{p(z+L_1)}, & z \leq -L_1, \\ \left[\cos h_2 L_2 + C_2 \sin h_1 L_1\right] e^{-q(z-L_2)}, & z \geq L_2. \end{cases} \tag{3}$$

Here $A$ is the normalized amplitude which can be calculated by integrating the longitudinal component of the Poynting vector's (Yariv and Yeh 2007). Assuming the value of the power per unit length along the $y$-axis to be of the order of 1 W/m (Yariv and Yeh 1984), we obtain the value of the normalized amplitude $A$ in Eq. (3) using the formula $A^2 = 8\pi/(c\int\limits_{-\infty}^{\infty} E_y H_z dz)$ for TE-modes and $A^2 = -8\pi/(c\int\limits_{-\infty}^{\infty} H_y E_z dz)$ for TM-modes, with $c$ being the speed of light in vacuum. The coefficients $C_1$ and $C_2$ in Eq. (3) are determined as:

$$C_1 = \frac{\delta p - h_1 \tan h_1 L_1 - \beta \nu}{\delta p \tan h_1 L_1 + h_1 - \beta \nu \tan h_1 L_1}, \quad C_2 = \frac{h_2 \tan h_2 L_2 - \sigma q}{h_2 + \sigma q \tan h_2 L_2}, \tag{4}$$

and the transverse components of the wave vector in each layer are defined as: $h_1^2 = k_0^2 \varepsilon_1 \mu_\perp - \beta^2$ and $h_2^2 = k_0^2 \varepsilon_{yy} \mu_2 - \beta^2$ for TE-modes, $h_1^2 = k_0^2 \mu_1 \varepsilon_\perp - \beta^2$ and $h_2^2 = k_0^2 \varepsilon_{xx} \mu_2 - (\varepsilon_{xx}/\varepsilon_{zz})\beta^2$ for TM-modes, and $p^2 = k_0^2 \varepsilon_3 - \beta^2$ and $q^2 = \beta^2 - k_0^2 \varepsilon_4$ for the substrate and cladding, respectively. Here $k_0 = \omega/c$ is the wave vector of the electromagnetic wave in air, $\sigma = \mu_2/\mu_4$, $\tau = \mu_1/\mu_2$, $\delta = \mu_1/\mu_3$, $\nu = \mu_a/\mu_1$ for TE-mode, and $\sigma = \varepsilon_{zz}/\varepsilon_4$, $\tau = \varepsilon_1/\varepsilon_{zz}$, $\delta = \varepsilon_1/\varepsilon_3$, $\nu = \varepsilon_a/\varepsilon_1$ for TM-mode, with $\mu_\perp = \mu_1 - \mu_a^2/\mu_1$ and $\varepsilon_\perp = \varepsilon_1 - \varepsilon_a^2/\varepsilon_1$ being the transverse permeability and permittivity, respectively. The dispersion equation, obtained from the boundary conditions consisting in continuity of the tangential field components at the boundaries of the media, can be written as

$$\left[ \delta\tau\, ph_2^2 + \sigma qh_1^2 + \beta\nu\left(\beta\nu\sigma q - \delta\sigma\, pq - \tau h_2^2\right)\right]\tan h_1 L_1 \cdot \tan h_2 L_2 +$$
$$h_2\left[h_1^2 - \delta\sigma\tau\, pq + \beta\nu\left(\beta\nu + \tau\sigma q - \delta\, p\right)\right]\tan h_1 L_1 + h_1\left(\tau h_2^2 - \delta\sigma\, pq\right)\tan h_2 L_2 - h_1 h_2\left(\delta\, p + \sigma\,\tau\, q\right) = 0 \ . \tag{5}$$

If $L_2 = 0$, the Eq. (5) transforms into a conventional dispersion relation for a three-layer waveguide structure (Adams 1981; Barnoski 1974).

The mode order is defined as the number of nodes of the profile function distribution within the waveguide which can be obtained from the condition $F_y(z) \to 0$. In order to fulfill this condition the profile functions in the waveguide layers can be written as:

$$\begin{cases} F_y\left(-L_1 \leq z \leq 0\right) = A \cdot \mathrm{sign}(C_1)\sqrt{1+C_1^2}\,\sin\left(h_1 z + \varphi_1\right), \\ F_y\left(0 \leq z \leq L_2\right) = A \cdot \mathrm{sign}(C_2)\sqrt{1+C_2^2}\,\sin\left(h_2 z + \varphi_2\right), \end{cases} \tag{6}$$

where $\varphi_1 = \arctan(1/C_1)$ and $\varphi_2 = \arctan(1/C_2)$ are the initial phases. Then the number of nodes in the chosen layer is:

$$\begin{cases} M_i = \left\{h_i L_i / \pi\right\}_{\mathrm{int}} + 1, \text{ if } \pi \cdot \left\{h_i L_i / \pi\right\}_{\mathrm{int}} - h_i L_i \leq \varphi_i \leq 0, \\ M_i = \left\{h_i L_i / \pi\right\}_{\mathrm{int}}, \text{ in all other cases.} \end{cases} \tag{7}$$

The subscript $i = 1, 2$ denotes the YIG and NC layers, respectively. The order of the mode is then determined as $M = M_1 + M_2$.

## 3. Numerical results and discussion

For the numerical analysis of the results obtained above, we take into account the dispersion of refractive indices (the Sellmeier's equations) of YIG, $SiO_2$, and GGG:

$$n_j^2 = 1 + \frac{A_1\lambda^2}{\lambda^2 - B_1^2} + \frac{A_2\lambda^2}{\lambda^2 - B_2^2} + \frac{A_3\lambda^2}{\lambda^2 - B_3^2} \ . \tag{8}$$

**Table 1.** Sellmeier coefficients in Eq. (8).

| Material | $A_1$ | $A_2$ | $A_3$ | $B_1$ (μm) | $B_2$ (μm) | $B_3$ (μm) |
|---|---|---|---|---|---|---|
| YIG (Johnson and Walton 1965) | 3.739 | 0.79 | - | 0.28 | 10 | - |
| GGG (Wood and Nassau 1990) | 1.7727 | 0.9767 | 4.9668 | 0.1567 | 0.01375 | 22.715 |
| $SiO_2$ (Malitson 1965) | 0.6961663 | 0.4079426 | 0.8974794 | 0.0684043 | 0.1162414 | 9.896161 |

The dispersion of $TiO_2$ is given by formula $n_{TiO_2}^2 = 5.913 + \dfrac{0.2441}{\lambda^2 - 0.0803}$ (Devore 1951).

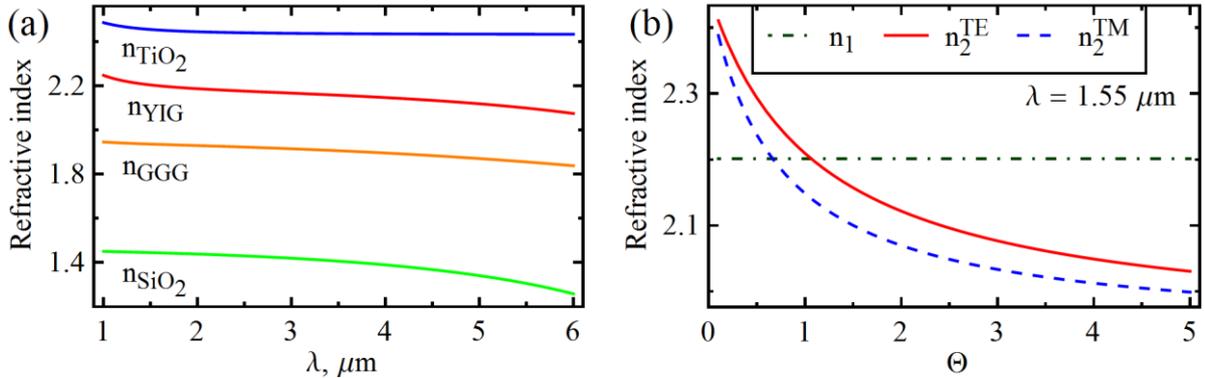

**Fig. 2.** Refractive indices dispersion $n(\lambda)$ for $TiO_2$, YIG, GGG and $SiO_2$ (a); the refractive indices of NC $n_2^{\mathrm{TE}} = \sqrt{\varepsilon_{yy}\mu_2}$ and $n_2^{\mathrm{TM}} = \sqrt{\varepsilon_{zz}\mu_2}$ as functions of $\Theta$ (b).

Figure 2(a) illustrates the dispersion of the refractive indices of the materials constituting the structure under consideration, calculated using the Sellmeier's equations within the range $\lambda = (1 \div 6)$ μm . For GGG, $TiO_2$ and $SiO_2$ layers the dielectric permittivities are: $\varepsilon_{GGG}(\lambda) = n_{GGG}^2(\lambda)$, $\varepsilon_{TiO_2}(\lambda) = n_{TiO_2}^2(\lambda)$, $\varepsilon_3(\lambda) = n_{SiO_2}^2(\lambda)$. The diagonal components of the dielectric permittivity tensor $\hat{\varepsilon}_{YIG}$ are assumed to be $\varepsilon_1 = n_{YIG}^2(\lambda)$ and those of the magnetic permeability tensor $\hat{\mu}_{YIG}$ can be taken as $\mu_1 = 1$ for the considered frequency regime (Gurevich and Melkov 1996). The off-diagonal material tensor components for YIG are $\varepsilon_a = -2.47 \cdot 10^{-4}$ and $\mu_a = 8.76 \cdot 10^{-5}$ (Torfeh and Le Gall 1981).

Figure 2(b) shows the refractive indices of the NC multilayer and YIG film obtained from the wave localization conditions $h_1 = 0$ and $h_2 = 0$. These conditions confine the waveguide propagation of the electromagnetic wave within either the YIG or the NC layer. Thus, one can introduce two propagation regimes: A-regime, in which the waveguide modes are guided by both YIG and NC layers, and B-regime for the modes guided by only one of these layers (Adams 1981). For the NC multilayer $n_2^{TE} = \sqrt{\varepsilon_{yy}\mu_2}$, $n_2^{TM} = \sqrt{\varepsilon_{zz}\mu_2}$. The refractive indices for TE- and TM-modes in the NC considerably differ ($n_2^{TE} > n_2^{TM}$), that is a consequence of its optical anisotropy, which, in turn, is due to the nanostructuring of the NC multilayer. The optical anisotropy of YIG, caused by its bigyrotropy, is much smaller, and in this connection its ~~the~~ refractive indices $n_1^{TE}$ and $n_1^{TM}$ are almost equal in the considered wavelength range ($n_1^{TE} \cong n_1^{TM} \equiv n_1 = \sqrt{\varepsilon_1\mu_1}$). The differences between $n_2^{TE}$, $n_2^{TM}$ and $n_1$ vary with $\Theta$, thus selecting the value of $\Theta$, one can set the appropriate waveguide regime for the chosen wavelength.

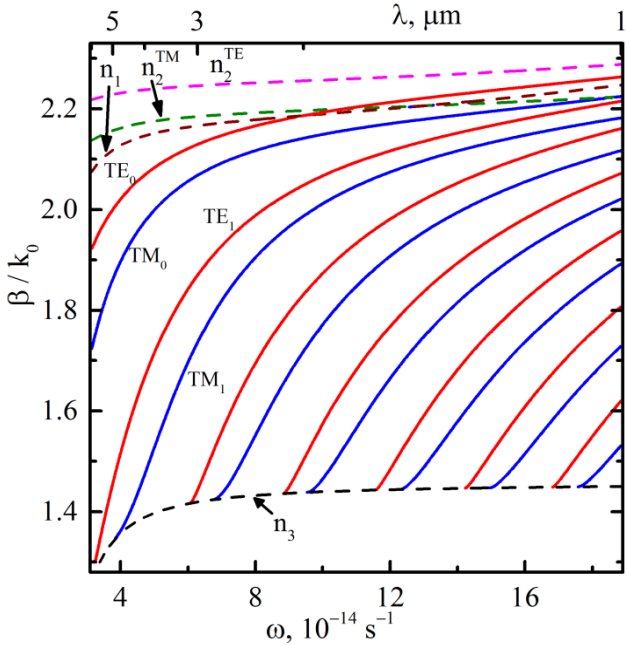

**Fig.3.** The spectra of TE- and TM-modes (normalized wave number $\beta/k_0$ vs angular frequency $\omega$) for the four-layer NC structure. The calculations are carried out for $L_1 = L_2 = 1$ μm and $\Theta = 0.65$ .

In Fig. 3 we present the spectra for TE- and TM-modes within the wavelength range $\lambda = (1 \div 6)$ μm. Each dispersion curve is characterized by the corresponding mode number $M = 0, 1, ...$ (as indicated in Fig. 3). The part of the $TE_0$-mode dispersion curve is located below $n_1$ ($\lambda > 2$ μm) and corresponds to the A-regime, while the other part is located above $n_1$ ($\lambda < 2$ μm) and corresponds to the B-regime of the waveguide propagation in the NC-layer. The dispersion curve of the $SiO_2$ substrate $n_3$ (the black dashed line) limits TE- and TM-modes from below, while the curve $n_2^{TE}$ (magenta dashed line) limits the modes from above. Also one can see that $n_2^{TE} > n_2^{TM}$ throughout the entire considered wavelength range, as it follows from the form of the NC permittivity tensor components and the permittivities of its constituent materials.

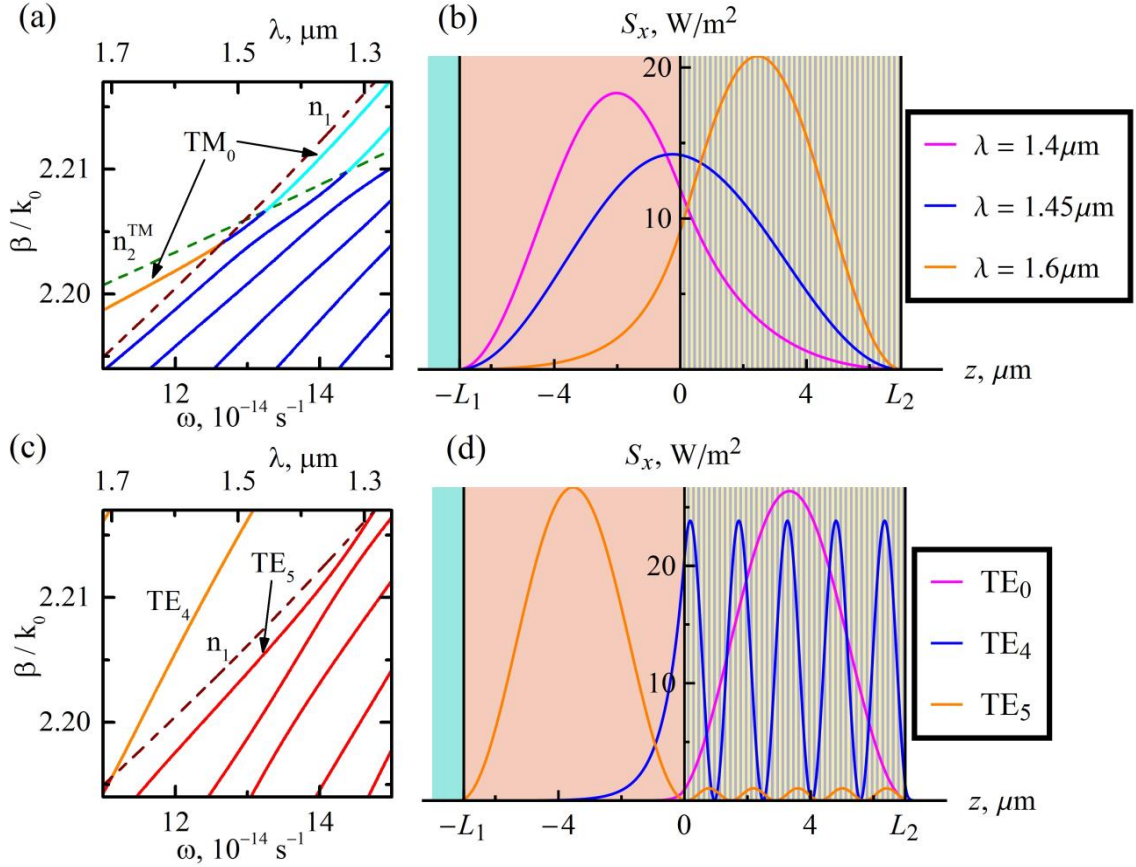

**Fig.4.** The spectra of TM- (a) and TE-modes (c) (normalized wave number $\beta / k_0$ vs angular frequency $\omega$) for the four-layer NC structure. The distribution of the longitudinal Poynting's vector component within the waveguide layers for the fundamental $TM_0$-mode (b) and for TE-modes (d) for $\lambda = 1.5\,\mu m$. The calculations are carried out for $L_1 = L_2 = 7\,\mu m$ and $\Theta = 0.65$.

In Fig. 4 the parameters $L_1$, $L_2$ and $\Theta$ are chosen so that the dispersion curves $n_2^{TM}$ and $n_1$ intersect near $\lambda = 1.5\,\mu m$. Thus from Fig. 4(a) one can see that the $TM_0$-mode transforms from the mode, guided by the layer 2 (orange line), to the mode, guided by the layer 1 (cyan line), passing through the area, where it propagates in the A-regime where the mode is guided by both layers (blue line). The peak of its Poynting's vector longitudinal component $S_x = -\dfrac{c}{8\pi} \mathrm{Re}(H_y E_z)$ shifts from the YIG-layer to the NC with the increase of $\lambda$ [Fig. 4(b)]. For the TE-polarization in this case the dispersion curve $n_2^{TE}$ lies higher, outside the depicted range of $\beta / k_0$, and several modes with the mode order up to $M = 4$ propagate in the B-regime in the NC-layer and, all the higher order TE-modes starting from the $TE_5$-mode are modes of the A-regime, as shown in Figs. 4(c) and 4(d).

## 4. Power flux analysis

In Fig. 5 the partial power flux redistributions of the TM-modes are demonstrated. The total power flux in the structure is $P = \displaystyle\int_{-\infty}^{\infty} S_x \, dz$. Thus, as the $TM_0$-mode changes the propagation regime with the increase of $\lambda$, the partial power flux in the YIG-layer $P_1 = \displaystyle\int_{-L_1}^{0} S_x \, dz / P$ decreases, while the part of the energy carried by the NC-layer $P_2 = \displaystyle\int_{0}^{L_2} S_x \, dz / P$ grows. Thus the main part of the energy flux in the structure moves from one waveguide layer to another one.

The partial power fluxes of the higher order modes (namely, the $TM_1$- and $TM_2$-mode) oscillate with the change of $\lambda$. As one can see from Fig. 5, the equal power flux distribution between two waveguide layers for $TM_0$-mode occurs at $\lambda \approx 1.45\,\mu m$, for $TM_1$-mode it takes place at $\lambda \approx 1.35\,\mu m$ and $\lambda \approx 1.45\,\mu m$, and for $TM_2$-mode the equal power fluxes are reached at the wavelengths $\lambda \approx 1.37\,\mu m$, $\lambda \approx 1.45\,\mu m$ and $\lambda \approx 1.6\,\mu m$, however in the latter case the oscillations of the power fluxes are less pronounced, and in the most of the wavelength range the power flux ratio between the guiding layers remains almost constant ($P_1 / P_2 \approx 1$).

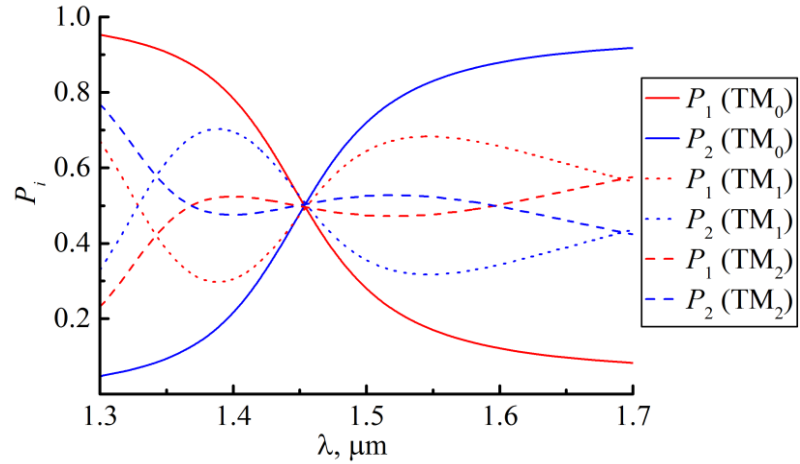

**Fig.5.** The TM-modes partial power fluxes $P_i$ as a function of the wavelength $\lambda$ (the subscripts $i = 1, 2$ denote the YIG and NC layers, respectively). The calculations are carried out for $L_1 = L_2 = 7$ μm and $\Theta = 0.65$.

According to our estimations, the fundamental $TM_0$-mode has the largest Q-factor $Q = 10 \log_{10}(P_1 / P_2)$ of the order of 6 dB, which can be achieved in the wavelength range of about 200 nm.

Therefore, a wavelength-tunable optical switch with the possibility of changing the logical state of the waveguide optical cell can be constructed on the base of the considered structure. Moreover, the structure can be used as a two-channel polarization splitter (at the fixed wavelength $\lambda$) where the $TE_0$-mode propagates in one layer, while the $TM_0$-mode propagates in another one.

## 5. Conclusion

The features of TE- and TM-polarized waves propagation in the four-layer nanocomposite-based magneto-optical waveguide structure are investigated in the long-wavelength approximation and the dispersion equation has been obtained considering the nanostructuring of the nanocomposite layer and taking into account bigyrotropy of the magnetic layer. Such a structure can be used as a polarization filter controlled with the geometrical parameters of the structure and with the so-called $\lambda$-tuning.

The distribution of the power flux across the structure demonstrates the ability either to spatially separate the different polarization modes within the structure or to allocate the main part of the power flux of the fundamental mode between the waveguide layers.

**Funding.** This research has received funding from the European Union's Horizon 2020 research and innovation program under the Marie Skłodowska-Curie grant agreement No. 644348 (N.N.D., Yu.S.D., M.K. and I.L.L.), MPNS COST Action MP1403 "Nanoscale Quantum Optics" (N.N.D., Yu.S.D., and I.L.L.), is supported by the grant from Ministry of Education and Science of Russian Federation: Project No. 14.Z50.31.0015 and No. 3.2202.2014/K (N.N.D., Yu.S.D., I.S.P. and D.G.S.) and also is supported by Ukrainian State Fund for Fundamental Research under project No. Φ71/73-2016 "Multifunctional Photonic Structures" (I.L.L.).